\documentclass{article}
\usepackage{epsfig}

\def\epsilon{\varepsilon}
\def\xmin{x_{\rm min}}\def\xmax{x_{\rm max}}
\def\vmin{v_{\rm min}}\def\gmin{\gamma_{\rm min}}

\title{Microscopic models of traveling wave equations}
\author{Eric \textsc{Brunet}\ and\ Bernard \textsc{Derrida}
   \\[4mm]
	\small Laboratoire de Physique Statistique,\\
         \small ENS, 24 rue Lhomond, 75005 Paris, France} 
\date{Computer Physics Communications \textbf{121--122} (1999) 376--381}

\begin{document}

\maketitle

\begin{abstract}
Reaction-diffusion problems are often described at a macroscopic
scale by partial derivative equations of the type
of the Fisher or Kolmo\-gorov-Petrovsky-Piscounov equation.
These equations have a continuous  family  of  front solutions, each of
them corresponding to a different velocity of the front.
By simulating systems of size up to $N=10^{16}$ particles at the
microscopic scale, where particles react and diffuse according to some
stochastic rules, we show that a single velocity is selected for the front.
This velocity converges logarithmically to the solution of the F-KPP
equation with minimal velocity when  the number~$N$ of particles increases.
A simple calculation of the effect introduced by the cutoff due to the
microscopic scale allows one to understand the origin of the logarithmic
correction.

\medbreak

\end{abstract}

\section{The Fisher equation}

The Fisher equation\cite{Fisher.Genes.37}, also called KPP equation (for
Kolmo\-gorov-Petrov\-sky-Piscounov\cite{KPP.Diffusion.37}) is widely used to
describe front propagation in many problems of physics, chemistry and
biology:
\begin{equation}
{\partial c \over \partial t} = {\partial^2c\over\partial x^2} +c-c^2.
\label{eqn:FK}
\end{equation}
Fisher first introduced this equation to represent ``The Wave of
Advance of Advantageous Genes''\cite{Fisher.Genes.37} in a population. The
concentration $c(x,t)$ was the fraction of individuals in a population at
position~$x$ and time~$t$ that exhibit some benefic genes,
and~(\ref{eqn:FK}) was used to describe how this favorable gene would
spread in the population. Equation~(\ref{eqn:FK}) can also model the
dynamics of sick individuals in a population during a viral contagious
infection, the proportion of burnt-out gases in a
combustion\cite{Kerstein.LatticeGas.86}, the
concentration of some species produced in a chemical reaction, etc.
It also appears in the mean field theory of directed
polymers in a random medium\cite{DerridaSpohn.Polymers.88} and in the
calculation of Lyapunov exponents of large sparse random
matrices\cite{CookDerrida.Lyapunov.90,vanZon.Lyapunov.98}.

In~(\ref{eqn:FK}), $c(x,t)$ represents a concentration, implying that,
for all~$x$ and~$t$:
\begin{equation}
0\le c(x,t)\le1.
\label{eqn:bornes}
\end{equation}

If we look at  solutions constant in space 
(${\partial c/\partial x}=0$),
equation~(\ref{eqn:FK}) becomes simply:
\begin{equation}
	{\partial c\over\partial t}=c-c^2.
\end{equation}
There are two stationnary solutions: $c=0$ (unstable fixed
point) and $c=1$ (stable fixed point). 

The role of the diffusion term ${\partial^2 c/\partial x^2}$
in~(\ref{eqn:FK}) is to spread any positive perturbation. Therefore, if
initially $c(x,0)>0$ in some region of space and $c(x,0)=0$ elsewhere, the
perturbation will grow, reach asymptotically the stable fixed point $c=1$
and spread throughout the whole space. Ultimately, the stable value $c=1$
will be reached everywhere.

To study how the stable region (where $c=1$) invades the unstable one
(where $c=0$), one can consider an initial condition where $c(x,0)$
decreases monotonically from $c(-\infty,0)=1$ to $c(+\infty,0)=0$. 
If at time $t=0$ the initial condition $c(x,0)$ decreases fast enough as
$x\to+\infty$ (in particular if $c(x,0)$ is a step function), the front
moves in the long time limit with a well defined
speed~$\vmin$\cite{Bramson.Convergence.83}: 
\begin{equation}
\vmin=2.
\end{equation}

To understand from~(\ref{eqn:FK}) why the velocity $\vmin=2$ is selected
looks a priori a very hard task. Equation~(\ref{eqn:FK}) is a
non-linear partial derivative equation and there is no way of writing the
full expression of $c(x,t)$ for an arbitrary initial condition. However,
the velocity $\vmin=2$ can be understood easily without solving the full
non-linear problem: let us assume that the front moves at some constant
speed~$v$. The concentration profile $c(x,t)$ takes then the form:
\begin{equation}
	c(x,t)=F_v(x-vt),
		\label{eqn:sol}
\end{equation}
where $F_v(z)$ is the solution of:
\begin{equation}
		F_v''+vF_v'+F_v-F_v^2=0,
		\label{eqn:eqdiff}
\end{equation}
with $F_v(-\infty)=1$ and $F_v(+\infty)=0$.
In the region where $F_v(z)\ll1$, {\it i.e.} far ahead of the front, one
can neglect in~(\ref{eqn:eqdiff}) the non-linear term. The general 
solution of the linearized
equation is a sum of two exponentials so that for~$z\to+\infty$ one of
these two exponentials dominates:
\begin{equation}
        F_v(z)\simeq A e^{-\gamma z},
	\label{eqn:decay}
\end{equation}
and, from the linearized version of~(\ref{eqn:eqdiff}), one finds that
$v$ and $\gamma$ are related by:
\begin{equation}
        v(\gamma)=\gamma+{1\over\gamma}.
        \label{eqn:v(g)}
\end{equation}
We see that the asymptotic decay~$\gamma$ in~(\ref{eqn:decay}) 
determines completely the velocity $v(\gamma)$ of the front.

What (\ref{eqn:v(g)}) tells us is that any velocity $v\ge \vmin$ ($\vmin=2$)
is possible for the front.
(It should be noted that $v<\vmin$ would also be possible
by allowing $\gamma$ to be complex. However, a front moving at such a speed
would take negative values in the tail, and this would violate
condition~(\ref{eqn:bornes}).)

The minimal speed $\vmin=2$ reached for~$\gmin=1$ has a special status:
it has been shown\cite{Bramson.Displacement.78,Jacob.Selection.85,%
vanSaarloos.MarginalStability.89,BrunetDerrida.Cutoff.97,%
Ebert.Relaxation.98} that if in its initial condition the front decays
faster than $e^{-\gmin x}$ (in particular, if~$c(x,0)$ is a step function),
then the front moves asymptotically with this minimal speed $\vmin$.
Moreover, in the long time limit, the position~$X(t)$ of the front is given
by
\begin{equation}
X(t)=2t-{3\over2}\ln t+O(1).
\label{eqn:transient}
\end{equation}
(In other words, the velocity of the front converges to $\vmin=2$ with a
leading correction\cite{Ebert.Relaxation.98} given by $-3/(2t)$. The
presence of a logarithmic correction in~(\ref{eqn:transient}) makes often a
precise determination of the asymptotic velocity difficult.)

\medbreak

Most properties of~(\ref{eqn:FK}) (selection of the minimal
velocity for steep enough initial conditions, logarithmic corrections to
the position as in~(\ref{eqn:transient})) can also be recovered
in a whole class\cite{vanSaarloos.MarginalStability.89} of front
equations where a stable region invades an unstable one. An example very
different from~(\ref{eqn:FK}) that we will consider below is:
\begin{equation}
c(x,t+1)= 1-\bigl[1-\int{\rm d}\alpha\,\rho(\alpha)c(x-\alpha,t) \bigr]^2,
\label{eqn:meanfield}
\end{equation}
where $\rho(\alpha)$ can be any density function~($\rho(\alpha)\ge0$ and
$\int{\rm d}\alpha\,\rho(\alpha)=1$). As for~(\ref{eqn:FK}), the uniform
solutions $c=0$ and $c=1$ are respectively unstable and stable and the
integral over~$\alpha$ spreads any positive perturbation as does the
diffusion term in~(\ref{eqn:FK}).

As for~(\ref{eqn:FK}), the linearized version of~(\ref{eqn:meanfield})
where terms quadratic in $c$ are neglected
determines the velocity. For an exponential decay~(\ref{eqn:decay})
of the front, one finds a dispersion relation
$v(\gamma)$ generalizing~(\ref{eqn:v(g)}):
\begin{equation}
v(\gamma)={1\over\gamma}\ln\left[2\int{\rm
d}\alpha\,\rho(\alpha)e^{\gamma\alpha}\right],
\label{eqn:dispersion}
\end{equation}
and for a steep enough initial condition the minimum velocity
\begin{equation}
\vmin=\min_\gamma v(\gamma)=v(\gmin)
\label{eqn:minv}
\end{equation}
is reached in the long time limit. The position~$X(t)$ is then
given\cite{BrunetDerrida.Cutoff.97} for large~$t$ by:
\begin{equation}
X(t)=\vmin t-{3\over2\gmin}\ln t+O(1).
\label{eqn:generaltransient}
\end{equation}
(Note that in general $\gmin$ in~(\ref{eqn:minv}) is finite except for
very particular choices of $\rho(\alpha)$.)

\section{The microscopic stochastic model}
\label{sec:stoch}

Front equations of type~(\ref{eqn:FK}) or~(\ref{eqn:meanfield})
originate often as the large-scale limit of microscopic stochastic
models\cite{Kerstein.LatticeGas.86,CookDerrida.Lyapunov.90,%
vanZon.Lyapunov.98,Kessler.Landscape.97,Breuer.FluctuationEffects.94,%
Breuer.MacroscopicLimit.95}. Here we study a particular microscopic
model which, as we will see, is described in the large scale limit by the
front equation~(\ref{eqn:meanfield}). We will compare the velocity measured
for this microscopic stochastic problem with the
velocity~(\ref{eqn:dispersion},\ref{eqn:minv}) expected for the traveling
wave equation~(\ref{eqn:meanfield}).

Our microscopic model\cite{BrunetDerrida.Cutoff.97} is defined as follows:
Imagine a population where each generation has exactly $N$ individuals.
Each individual~$i$ ($1\le i\le N$) at generation~$t$ ($t$ is an integer)
is characterized by its fitness $x_i(t)$, a real number
representing its adaptation to the environment. The state of the system at
any time~$t$ is completely determined by the $N$ numbers $x_i(t)$.

At time $t=0$, we set  $x_i(0)=0$ for all~$i$ (but this choice of
initial condition is actually unimportant in the long time limit).
By definition of the model, the $x_i(t)$ evolve from 
generation~$t$ to generation~$t+1$ with the following rule:
\begin{equation}
x_i(t+1)=\max\left[x_{m_i}(t)+\alpha_i,\,x_{f_i}(t)+\alpha'_i\right],
\label{eqn:evolution}
\end{equation}
where $m_i$ and $f_i$ are the two parents of the new individual~$i$, 
chosen at random in the previous generation~$t$ (in other words, $m_i$ and
$f_i$ are random integers uniformly distributed between~$1$ and~$N$), and
where $\alpha_i$ and $\alpha'_i$ are random numbers independently chosen
according to some probability distribution~$\rho(\alpha)$ representing
random mutations. So
at each generation,
the $m_i$, $f_i$, $\alpha_i$ and $\alpha'_i$ are independent and new
values are chosen at every time step.

Under the dynamics~(\ref{eqn:evolution}), the cloud of~$N$ points $x_i(t)$
moves along the line and we want to determine its asymptotic
velocity~$v_N$, that is:
\begin{equation}
v_N=\lim_{t\to+\infty} {X(t)\over t},
\end{equation}
where
\begin{equation}
X(t)={1\over N}\sum_{i=1}^N x_i(t).
\end{equation}

Let us now see how one can relate this microscopic model to the traveling
wave equation~(\ref{eqn:meanfield}). We define $c(x,t)$ as
the fraction of population which has a fitness
larger than~$x$:
\begin{equation}
c(x,t)={1\over N}\sum_{i=1}^N \Theta\big(x_i(t)-x\big).
\label{eqn:frdef}
\end{equation}
(By convention, we choose here $\Theta(x)=1$ if $x>0$ and $\Theta(x)=0$ if
$x\le0$.) Obviously, $c(x,t)$ is a monotonic decreasing function of $x$
going from $c(-\infty,t)=1$ to $c(+\infty,t)=0$. At time $t=0$, we have
$c(x,0)=1$ for $x<0$ and $c(x,0)=0$ for $x\ge0$, so the initial condition
is a step function. It should be noted that~$c(x,t)$ can only take values
which are integral multiples of $1/N$. 

Clearly from~(\ref{eqn:frdef}), the position~$X(t)$ of the cloud of points
can be rewritten with the function $c(x,t)$ as:
\begin{equation}
  X(t)=X(0)+\int_{-\infty}^{+\infty}{\rm d}x\,\left[c(x,t)-c(x,0)\right].
\end{equation}

   Given the positions $x_i(t)$ of all the particles (or, equivalently¸
given the function $c(x,t)$), the $x_i(t+1)$ obtained
from~(\ref{eqn:evolution}) are independent random
variables. Therefore, if we fix $c(x,t)$, the average $\bigl\langle
c(x,t+1)\bigr\rangle$ over the dynamics
between time~$t$ and~$t+1$ gives:
\begin{equation}
\bigl\langle c(x,t+1)\bigr\rangle
= 1-\Bigl[1-\int{\rm d}\alpha\,\rho(\alpha)c(x-\alpha,t) \Bigr]^2,
\label{eqn:stoch}
\end{equation}
which, except for the~$\langle\ \rangle$, is exactly~(\ref{eqn:meanfield}).
However, if we try to average over the whole history ({\it i.e.} over all
the timesteps), we need to average terms quadratic in~$c$ on the right hand
side of~(\ref{eqn:stoch}). This means $\langle c(x,t+1)\rangle$ is not only
related to $\langle c(x,t)\rangle$, but also to correlations like $\langle
c(x,t)c(x',t)\rangle$, and this makes the problem very difficult for
finite~$N$.  On the other hand, if we neglect these correlations (and one
can argue that these correlations are small for large enough~$N$) and
replace $\langle c(x,t)c(x',t)\rangle$ with $\langle c(x,t)\rangle\langle
c(x',t)\rangle$, then~(\ref{eqn:stoch}) reduces exactly
to~(\ref{eqn:meanfield}).  So~(\ref{eqn:meanfield}) can be thought as the
mean-field (or large~$N$) version of the microscopic 
model~(\ref{eqn:evolution}).

As the initial condition $c(x,0)$ given by~(\ref{eqn:frdef}) is a step
function, the mean field equation predicts a front moving at the minimum
velocity~$\vmin$ given by~(\ref{eqn:dispersion},~\ref{eqn:minv}).

\section{Direct simulations}
\label{sec:low}

We have simulated the microscopic model~(\ref{eqn:evolution}) for several
choices of the distribution $\rho(\alpha)$: the uniform distribution in the
range~$0\le\alpha\le1$ 
\begin{equation}
\rho_{\rm uni}(\alpha)=\Theta(\alpha)\Theta(1-\alpha),
\label{eqn:rhouni}
\end{equation}
the exponential distribution
\begin{equation}
\rho_{\rm exp}(\alpha)=\Theta(\alpha)e^{-\alpha},
\end{equation} 
and a discrete distribution
\begin{equation}
\rho_{\rm disc}(\alpha)=0.25\,\delta(1-\alpha)+0.75\,\delta(\alpha).
\end{equation}
We have also simulated a generalization of the problem (Martian genetics)
where each new individual at time~$t+1$ has {\em three} parents, so
that~(\ref{eqn:evolution}) is replaced by the max over three terms, with
the effect of mutations distributed according to~(\ref{eqn:rhouni}).

The minimal value of the speed~$\vmin$ and the corresponding decay
rate $\gmin$ of the deterministic front equations~(\ref{eqn:meanfield}, 
\ref{eqn:dispersion}, \ref{eqn:minv}) for 
these four cases are given in table~\ref{tab:v}.

\begin{table}[ht] 
\begin{tabular}{lllll}
\hline
model		& $\rho=\rho_{\rm uni}$ 	& $\rho=\rho_{\rm exp}$ 
		& $\rho=\rho_{\rm disc}$  	& Martian model\\
\hline
\hline
$\gmin$ 	& 5.262\,076\ldots 		& 0.626\,635\ldots 
		& 2.553\,245\ldots 		& 8.133\,004\ldots\\
$\vmin$		& 0.815\,172\ldots 		& 2.678\,347\ldots
		& 0.810\,710\ldots		& 0.877\,338\ldots \\
\hline
\end{tabular}
\caption{Values of $\gmin$ and $\vmin$ for different models.}
\label{tab:v}
\end{table}

\begin{figure}[ht]
\centerline{\epsfig{file=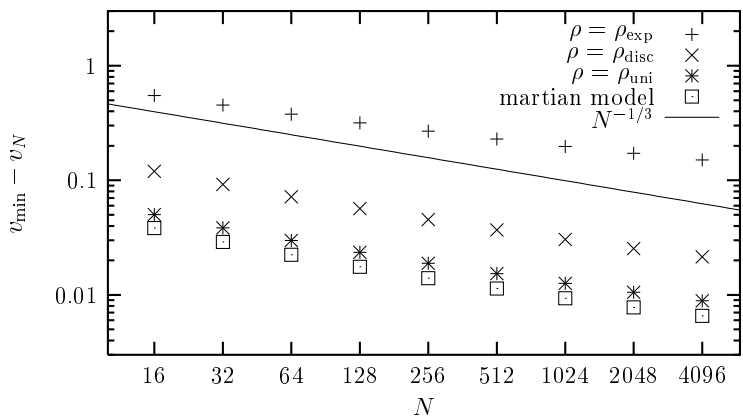,width=0.8\linewidth}}
\caption{Log-log plot of the difference between the speed~$\vmin$ given by
the mean-field theory and the speed~$v_N$ measured in Monte-Carlo
simulations for the four models as a function of the number~$N$ of
particles. The straight line represents $N^{-1/3}$.}
\label{fig:lowN}
\end{figure}

For these four models, we have simulated~(\ref{eqn:evolution}) for $T=10^7$
timesteps after a transient time of $T'=10^6$ timesteps to eliminate the
effect of initial conditions. For several choices of $N$, we have measured
the speed as:
\begin{equation}
v_N={X(T+T')-X(T')\over T}.
\label{eqn:T}
\end{equation}
Figure~\ref{fig:lowN} represents the difference between the mean-field
speed~$\vmin$ (given in table~\ref{tab:v}) and the speed~$v_N$ measured in
the simulation for several choices of~$N$ (16, 32, 64, 128, 256, 512, 1024,
2048 and 4096). 

We see on figure~\ref{fig:lowN} that a single speed $v_N$ is selected in
the microscopic model. This speed is always lower than the speed~$\vmin$
of the deterministic front equation, and the difference $\vmin-v_N$ seems
to decay like~$N^{-1/3}$ for all the variants of the model. The effective
power law exponent seems however to decrease slowly as $N$ increases.

In order to confirm the $N^{-1/3}$ decay of figure~(\ref{fig:lowN}), we
tried to increase~$N$, but as in the direct simulation computer time scales
like $N\times T$, it is very hard to make $N$ much larger than $10^4$ or
$10^5$ for a number  $T=10^7$ of timesteps.

\section{Highly parallel simulations}
\label{sec:high}

We are now going to describe a computational trick which we
developped\cite{BrunetDerrida.Cutoff.97} for some particular
distributions~$\rho(\alpha)$ such as 
\begin{equation}
\rho(\alpha)= p \delta(\alpha-1) + (1-p) \delta(\alpha),
\label{eqn:para}
\end{equation}
allowing to simulate the microscopic model for a huge number of points ($N$
up to $10^{16}$).  We restrict $p$ to the range $0<p<0.5$ (to avoid one of
the rare cases where~(\ref{eqn:dispersion}) has no minimum for a 
finite~$\gmin$).

For the distribution~(\ref{eqn:para}), the~$x_i(t)$ are always integers if
they are so at $t=0$
and the concentration~$c(x,t)$ as defined by~(\ref{eqn:frdef}) is
constant between any pair of consecutive integers. 
We call respectively $\xmin$ and~$\xmax$ the
positions of the leftmost and rightmost particles at time~$t$
and~$w=\xmax-\xmin+1$ the width of the front.
We observed in our simulations that $w$ is typically of order $\ln N$, so
that even for~$N$ as huge as $10^{16}$, the number of possible values of
the $x_i(t)$ at a given time is very limited, and the whole information in
$c(x,t)$ is carried by the number of particles at each integer~$x$ 
between~$\xmin$ and~$\xmax$.

Knowing the function~$c(x,t)$ at time~$t$, we generate~$c(x,t+1)$.
As at time $t$, all the $x_i(t)$ satisfy $\xmin\le x_i(t)\le\xmax$,
from~(\ref{eqn:evolution}) and~(\ref{eqn:para})
the positions~$x_i(t+1)$ will lie between~$\xmin$ and~$\xmax+1$. The
probability~$p_k$ that a given particle~$i$ will be located at
position~$\xmin+k$ at time~$t+1$ is
\begin{equation}
  p_k=\bigl\langle c(\xmin+k-1,t+1)\bigr\rangle-\bigl\langle
  c(\xmin+k,t+1)\bigr\rangle,
\end{equation}
with $\bigl\langle c(x,t+1)\bigr\rangle$ given
by~(\ref{eqn:stoch}). Obviously, $p_k\neq0$ only for $0 \le k \le w$.
The probability to have, for every $k$, $n_k$ particles at location
 $\xmin+k$ at time~$t+1$ is given by
\begin{eqnarray}
P(n_0,n_1,\ldots,n_w)=&&{N!\over n_0!\,n_1!\,\ldots\,n_w!}\
      p_0^{n_0}\,p_1^{n_1}\,\ldots\,p_l^{n_w}\nonumber\\
      &&\times\delta(N-n_0-n_1-\cdots-n_w).
\label{eqn:proba}
\end{eqnarray}
Using a random number generator for a binomial
distribution\cite{NumericalRecipes}, expression~(\ref{eqn:proba}) allows us to
generate the random numbers~$n_k$ directly\cite{BrunetDerrida.Cutoff.97}. 

\begin{figure}[ht]
\centerline{\epsfig{file=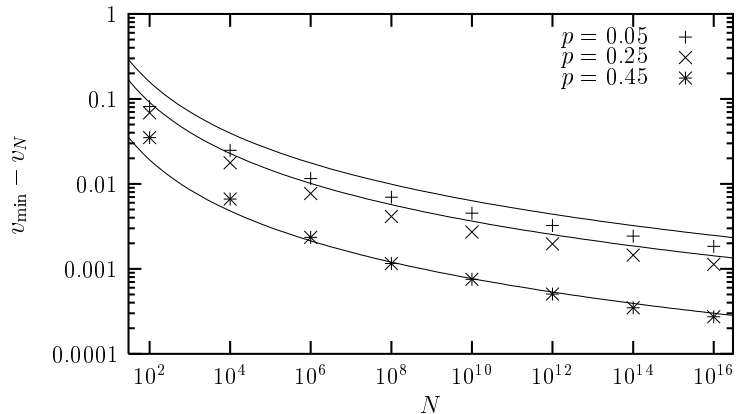,width=0.8\linewidth}}
\caption{Differences between the speed obtained by the mean-field theory and
the speed measured in highly parallel simulations for different values 
of~$p$ as a function of the number~$N$ of particles. The lines represent
for each value of $p$ the prediction~(\ref{eqn:result}).}
\label{fig:highN}
\end{figure}

   We have measured from~(\ref{eqn:T}) the velocity~$v_N$ of the front for
several choices of~$p$ (0.05, 0.25 and~0.45) and for $N$ ranging from~100
to~$10^{16}$. Figure~\ref{fig:highN} shows the results together with 
functions $A(p)/\ln^2N$ given in the next section.

Clearly the apparent power law of figure~\ref{fig:lowN} does not persist as
$N$ increases and the simulations indicate that $\vmin-v_N\sim\ln^{-2}N$
for large~$N$.  We are going to see in the next section that this
logarithmic correction has a simple origin.

\section{Effect of the cutoff}
\label{sec:cutoff}

The two main differences between the traveling wave
equation~(\ref{eqn:meanfield}) and the microscopic
model~(\ref{eqn:evolution}) is that the microscopic model is stochastic and
that $c(x,t)$ varies by steps multiple of $1/N$ (see (\ref{eqn:frdef})).

\begin{figure}[ht]
\centerline{\epsfig{file=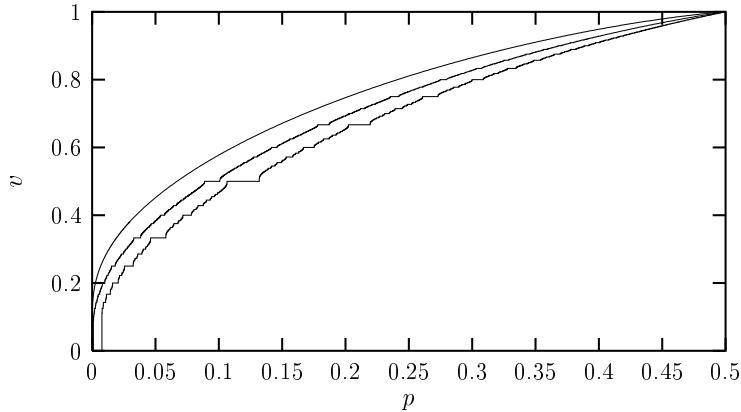,width=0.8\linewidth}}
\caption{Comparaison of the velocity $\vmin$ valid for infinite~$N$~(top curve)
with the velocity of the model with a cutoff $1/N$ for $N=512$~(middle curve) 
and $N=64$~(bottom curve) as a function of $p$.}
\label{fig:step}
\end{figure}

The effect of the noise is hard to treat analytically and we have not
succeeded yet to make a satisfactory theory of it.  The effect of
discretization can be however understood rather simply: let us
modify~(\ref{eqn:meanfield}) by imposing, after each timestep,
$c(x,t+1)=0$ if the value given by~(\ref{eqn:meanfield}) is smaller than
$1/N$, (in other words, we put a cutoff in the deterministic model to mimic
the fact that $c(x,t)$ changes by steps of $1/N$). The velocity $v_N'$ of
this deterministic model with a cutoff can be easily measured as under
these dynamics the front reaches rapidly a periodic
regime\cite{BrunetDerrida.Cutoff.97}. As the cutoff goes to zero
(\emph{i.e.} as $N$ goes to infinity), the speed $v_N'$ converges to the
mean-field speed~$\vmin$ given by (\ref{eqn:dispersion})
and~(\ref{eqn:minv}).  On figure~\ref{fig:step} we compare this speed for
$N=64$ and $N=512$ with $\vmin$ for $\rho(\alpha)$ given
by~(\ref{eqn:para}) with $0\le p\le 0.5$. (One can note that  the speed
gets locked to rational values as $p$ varies.)

The speed $v'_N$ of this new deterministic model can be calculated
an\-a\-lyt\-i\-cally\cite{BrunetDerrida.Cutoff.97} for large~$N$: 
\begin{equation}
v_N' \simeq \vmin -{\pi^2\gmin^2v''(\gmin)\over 2\ln ^2 N},
\label{eqn:result}
\end{equation}
where $v(\gamma)$ is given by~(\ref{eqn:dispersion}).
Comparing the prediction~(\ref{eqn:result}) with the results of 
the simulations in
figure~\ref{fig:highN} gives a good, though not perfect, agreement.
This indicates that the slow convergence of the velocity of the stochastic
model is controled by the effect of the cutoff.

\section{Conclusion}

We have seen that, for a very particular microscopic stochastic
model~(\ref{eqn:evolution}), it was possible to simulate systems of
$10^{16}$ particles. Our model is so particular that there is no hope that
our trick could be extended to large classes of statistical physics
problems. In our case, however, going to very large~$N$ enabled us to
clearly discriminate between a power law and a logarithmic correction.

The fact that the microscopic scale selects a single
velocity\cite{BrunetDerrida.Cutoff.97,Paquette.Renormalization.94,%
Mai.Ordering.97} with a 
logarithmic correction due to a cutoff seems to appear in several related
problems (reaction-diffusion\cite{Kerstein.LatticeGas.86,%
Kessler.Stability.98}, kinetic theory\cite{vanZon.Lyapunov.98}). Even
model~(\ref{eqn:evolution}) can be introduced in many different contexts,
like directed polymers in a random medium (where $x_i(t)$ would be the free
energy of a polymer of length~$t$ ending at position~$i$), or growth
problems (where $x_i(t)$ would be the height variables). Still, we do not
know yet whether the prediction~(\ref{eqn:result}) based simply on the effect 
of the cutoff gives the exact large-$N$ behavior of the
model~(\ref{eqn:evolution}), or whether a more sophisticated theory is
needed to explain the results of section~\ref{sec:cutoff}.

\bigskip

We thank M. Droz for communicating us several relevant references.

\end{document}